\documentclass{article}
\usepackage{icmc2020template}
\usepackage{times}
\usepackage{ifpdf}
\usepackage{soul}
\usepackage[english]{babel}
\usepackage{subfig}
\usepackage{booktabs}
\usepackage{listings}

\usepackage{amsmath} 
\usepackage{amssymb} 
\usepackage{amsfonts}

\def\papertitle{OrchideaSOL: a dataset of extended instrumental techniques for computer-aided orchestration}
\def\firstauthor{Carmine Emanuele Cella}
\def\secondauthor{Daniele Ghisi}
\def\thirdauthor{Vincent Lostanlen}
\def\fourthauthor{Fabien L\'evy}
\def\fifthauthor{Joshua Fineberg}
\def\sixthauthor{Yan Maresz}

\newcommand{\kk}{\kern-0.07em}
\newcommand{\muskern}{\kern-.15ex } 

\newcommand{\mf}{\textnormal{\textbf{\textit{m\muskern f}}}}
\newcommand{\mezzopiano}{\textnormal{\textbf{\textit{m\muskern p}}}}
\newcommand{\p}{\textnormal{\textbf{\textit{p}}}}
\newcommand{\pp}{\textnormal{\textbf{\textit{p\muskern p}}}}
\newcommand{\f}{\textnormal{\textbf{\textit{f}}}}
\newcommand{\ff}{\textnormal{\textbf{\textit{f\muskern f}}}}

\newenvironment{sloppypar*}
 {\sloppy\ignorespaces}
 {\par}

\makeatletter
\newcommand\dynmark[1]{{\normalfont\bfseries\itshape
  \@tfor\next:=#1\do{\put@muskern\next}\/}}
\newcommand{\put@muskern}{\let\put@muskern\muskern}
\makeatother

\newif\ifpdf
\ifx\pdfoutput\relax
\else
   \ifcase\pdfoutput
      \pdffalse
   \else
      \pdftrue
  \fi
\fi

\ifpdf 
  \usepackage[pdftex,
    pdftitle={\papertitle},
    pdfauthor={\firstauthor, \secondauthor, \thirdauthor},
    bookmarksnumbered, 
    pdfstartview=XYZ 
   ]{hyperref}

  \usepackage[pdftex]{graphicx}
  \DeclareGraphicsExtensions{.pdf,.jpeg,.png}

  \usepackage[figure,table]{hypcap}

\else 
  \usepackage[dvips,
    bookmarksnumbered, 
    pdfstartview=XYZ 
  ]{hyperref}  

  \usepackage[dvips]{epsfig,graphicx}
  \DeclareGraphicsExtensions{.eps}
  \usepackage[figure,table]{hypcap}
\fi

\hypersetup{
    colorlinks,%
    citecolor=black,%
    filecolor=black,%
    linkcolor=black,%
    urlcolor=black
}

\title{\papertitle}

 \sixauthors
   {0.5in}
   {\firstauthor} {University of California, Berkeley \\ CNMAT \\ %
     {\tt \href{carmine.cella@berkeley.edu}{carmine.cella@berkeley.edu}}}
   {\secondauthor} {University of California, Berkeley \\ CNMAT \\ %
     {\tt \href{danieleghisi@berkeley.edu}{danieleghisi@berkeley.edu}}}
   {\thirdauthor} { New York University\\ MARL \\ 
     {\tt \href{mailto:vl1019@nyu.edu}{vl1019@nyu.edu}}}
   {\fourthauthor} {University of Music and Theatre, Leipzig}
     {\fifthauthor} {Boston University}
   {\sixthauthor} {Conservatoire de Paris}

\begin{document}
\capstartfalse
\maketitle
\capstarttrue

\begin{abstract}
This paper introduces OrchideaSOL, a free dataset of samples of extended instrumental playing techniques, designed to be used as default dataset for the \emph{Orchidea} framework for target-based computer-aided orchestration. 

OrchideaSOL is a reduced and modified subset of \emph{Studio On Line}, or SOL for short, a dataset developed at Ircam between 1996 and 1998. We motivate the reasons behind OrchideaSOL and describe the differences between the original SOL and our dataset. We will also show the work done in improving the dynamic ranges of orchestral families and other aspects of the data.

\end{abstract}

\section{Introduction}\label{sec:introduction}

\begin{figure}[h]
\centering
\includegraphics[width=\columnwidth]{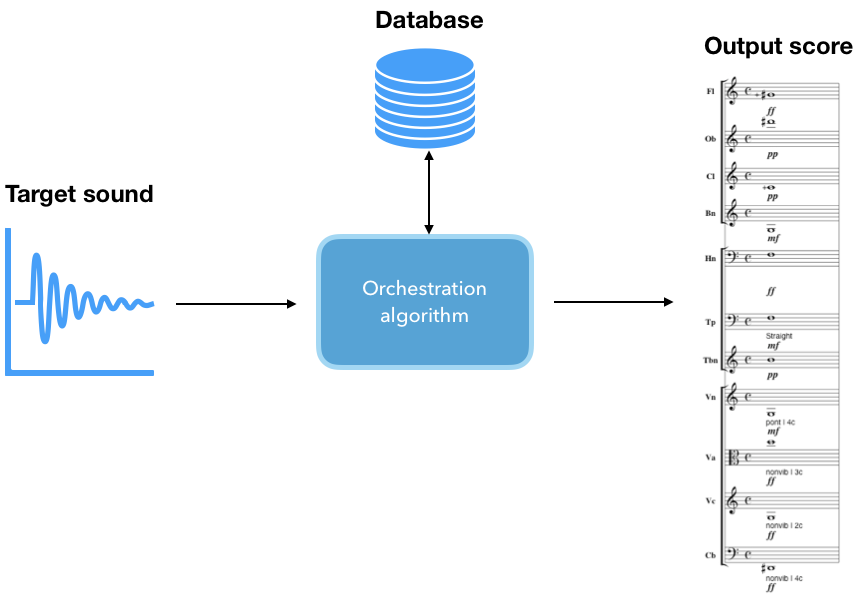}
\caption{A diagram of a typical system for computer-aided orchestration. \label{fig:aided_orch}}
\end{figure}

Target-based computer-aided orchestration is a set of techniques that help composers find combinations of orchestral sounds matching a given sound \cite{maresz2013computer}.
Typically, computer-aided orchestration systems consist of algorithms that compute a large number of combinations of audio samples in a dataset, corresponding to instrumental notes, trying to find the one combination that is ``more similar'' to the target (with respect to some metrics).
Solutions to this problem are proposed as orchestral scores, often ranked by similarity between the target sound and the mixture of audio samples in the dataset.
Figure \ref{fig:aided_orch} shows a typical workflow of a computer-aided orchestration system.

In such a context, the quality of the proposed orchestrations heavily depends on the quality and on the overall consistency of the sounds in the dataset.
An essential aspect of this consistency is given by the dynamic ranges across instrumental families.
Suppose, for example, that a flute note that is tagged as {\pp} has a waveform amplitude greater than a trumpet note tagged as {\ff}.
In this case, the orchestration solutions would work poorly in real-life orchestral scenarios.

Among the state-of-the-art systems for computer-aided orchestration there is the \emph{Orch*} family: \emph{Orchid\'ee} (2008) \cite{carpentier2010interacting}, \emph{Orchids} (2013) \cite{caetano2020}, \emph{Orchidea} (2017) \cite{gillick2019estimating}, developed at Ircam (Paris) and at the Haute \'Ecole de Musique (Geneva).
In this paper, we focus on the last iteration, Orchidea\footnote{For more information on the Orchidea system for computer-aided orchestration, please visit: \url{http://www.orch-idea.org}}.

\begin{sloppypar}
Until now, Orchidea relied on the Studio On Line (SOL) dataset \cite{ballet1999studio}, in a specific distribution, called 0.9.2, originally created as default dataset for Orchids.
This version differs from the original SOL in several regards. 
First, the amplitude level of each audio sample has been normalized: this means that the genuine dynamic ranges of each instrument family have been lost.
Moreover, the licensing of the dataset is not free and requires a Premium account on the Ircam forum\footnote{See https://forum.ircam.fr/ for more information.}.
\end{sloppypar}

The effects generated by the normalization of the files severely impacted the results of Orchidea. Thus, we decided to recover and process the original recordings of SOL to create three new datasets, one of them being OrchideaSOL. 
The three versions have different sizes and features and are distributed under different licenses:
\begin{itemize}
    \item \emph{TinySOL}: a small subset of SOL including only ordinario sounds, free for non-commercial usage;
    \item \emph{OrchideaSOL}: the version documented in this paper, featuring a certain number of extended techniques (playing techniques), free for non commercial usages after subscription to the Ircam Forum;
    \item \emph{FullSOL}: the full set of samples originally present in SOL, requiring a Premium account on the Ircam forum to be used.
\end{itemize}{}

Although in this paper we will specifically focus on OrchideaSOL, all three vesions have been processed in the same way, in order to correct and improve the dynamic ranges and other minor problems.

\section{Studio On Line: a brief history}
\emph{Studio On Line} \cite{ballet1999studio} is a dataset of instrumental sounds featuring a rich set of extended techniques commonly found throughout 20th century Western music.
SOL was recorded in the \emph{Espace de projection} of Ircam between 1996 and 1998 \cite{sollevyreport}.
Over the past two decades, SOL has served as a reference dataset for many Ircam projects.
The project also included software for sound processing and transformation.
The head of the project was Guillaume Ballet, while the artistic managers were Joshua Fineberg (in 1997) and Fabien L\'evy (in 1998).

The recordings were carried out in two phases.
The first phase included the recordings of Flute, Oboe, Clarinet, Bassoon, Horn, Trumpet, Trombone, Violin, Viola, Cello and Double Bass.
The second phase was planned to include more instruments and doublings \cite{solfinebergreport}; yet, only Tuba, Harp, Guitar, Alto Sax and Accordion were eventually accomplished.

\begin{figure}[h]
\centering
\includegraphics[width=6cm]{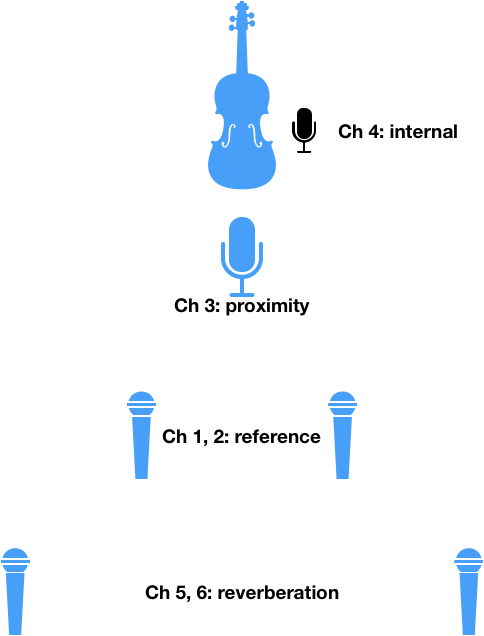}
\caption{Spatial arrangement of the microphones during the recording sessions of SOL. \label{fig:mikes}}
\end{figure}

All instruments were recorded in a six-channel format (see Fig. \ref{fig:mikes}):
\begin{itemize}
    \item a stereo couple (track 1 and 2) was used as a reference signal;
    \item a proximity microphone (track 3), with minimal reverberation, was used to record the signal at a fairly high level even for very soft sounds;
    \item a so-called `internal' microphone (track 4), either an aerial microphone placed inside the instrument or a contact microphone, was used to record sounds mostly for the purpose of acoustics studies;
    \item two bi-directional microphones with a figure-8 pattern (tracks 5 and 6), placed far from the musician, were used to capture the reverberation.
\end{itemize}

\begin{sloppypar}
More than 125k samples (sampled at 48 kHz with resolution of 24 bit) were recorded \cite{sollevyreport}: ordinario sounds were sampled at least at three levels of dynamics (usually \pp, \mf and \ff) and with a semitonal resolution. Woodwind instruments also included quarter tones. Most of the other playing techniques were sampled with a coarser resolution in pitch and/or in dynamics.
\end{sloppypar}

\section{OrchideaSOL}
We have built OrchideaSOL starting from the original version of SOL, properly cleaned and trimmed, resampled at 44.1kHz (with a bit depth of 24 bits), and made monophonic by only keeping the third channel (proximity microphone).
The rationale behind this choice was to avoid the normalization introduced in later versions of the base, such as the one used by Orchids, as well as some resampling issues.
The proximity microphone provided the cleanest audio recording quality.
We have subsequently performed the operation described in the following subsections.

\begin{figure*}[h]
\centering
\subfloat[\label{fsol}]{%
      \includegraphics[width=0.33\textwidth]{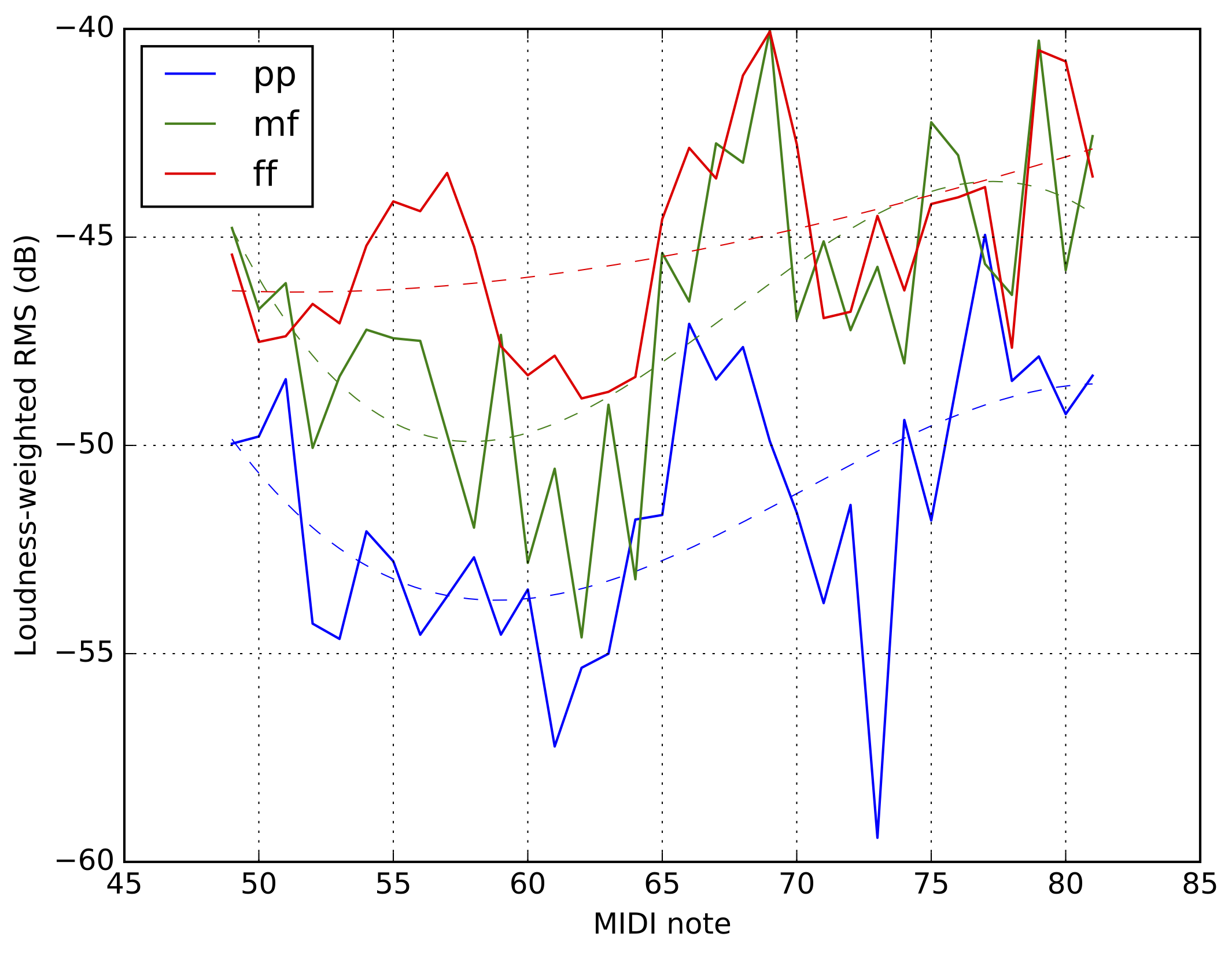}}
\hspace{\fill}
   \subfloat[\label{fsol09} ]{%
      \includegraphics[width=0.33\textwidth]{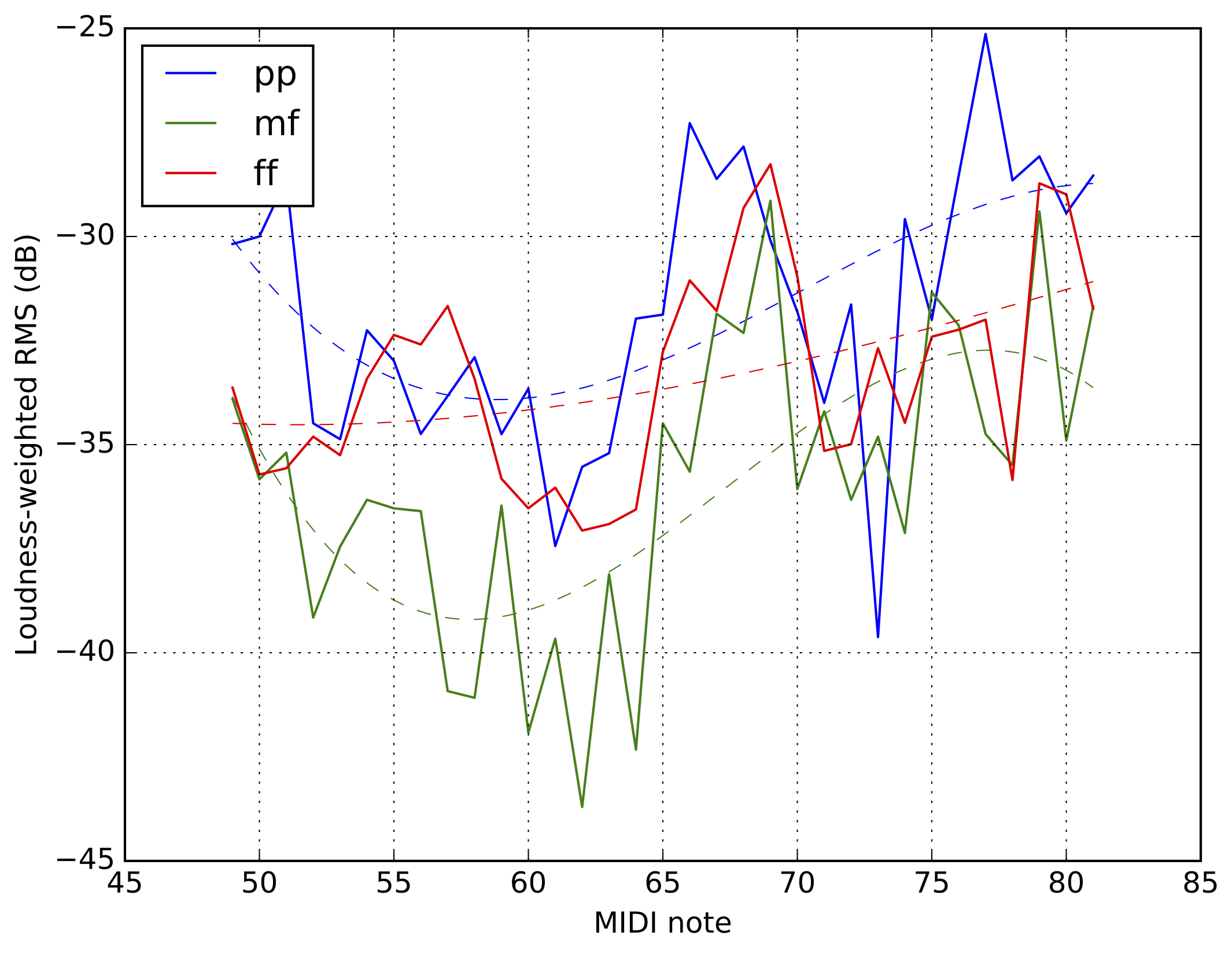}}
\hspace{\fill}
   \subfloat[\label{forchideasol}]{%
      \includegraphics[width=0.33\textwidth]{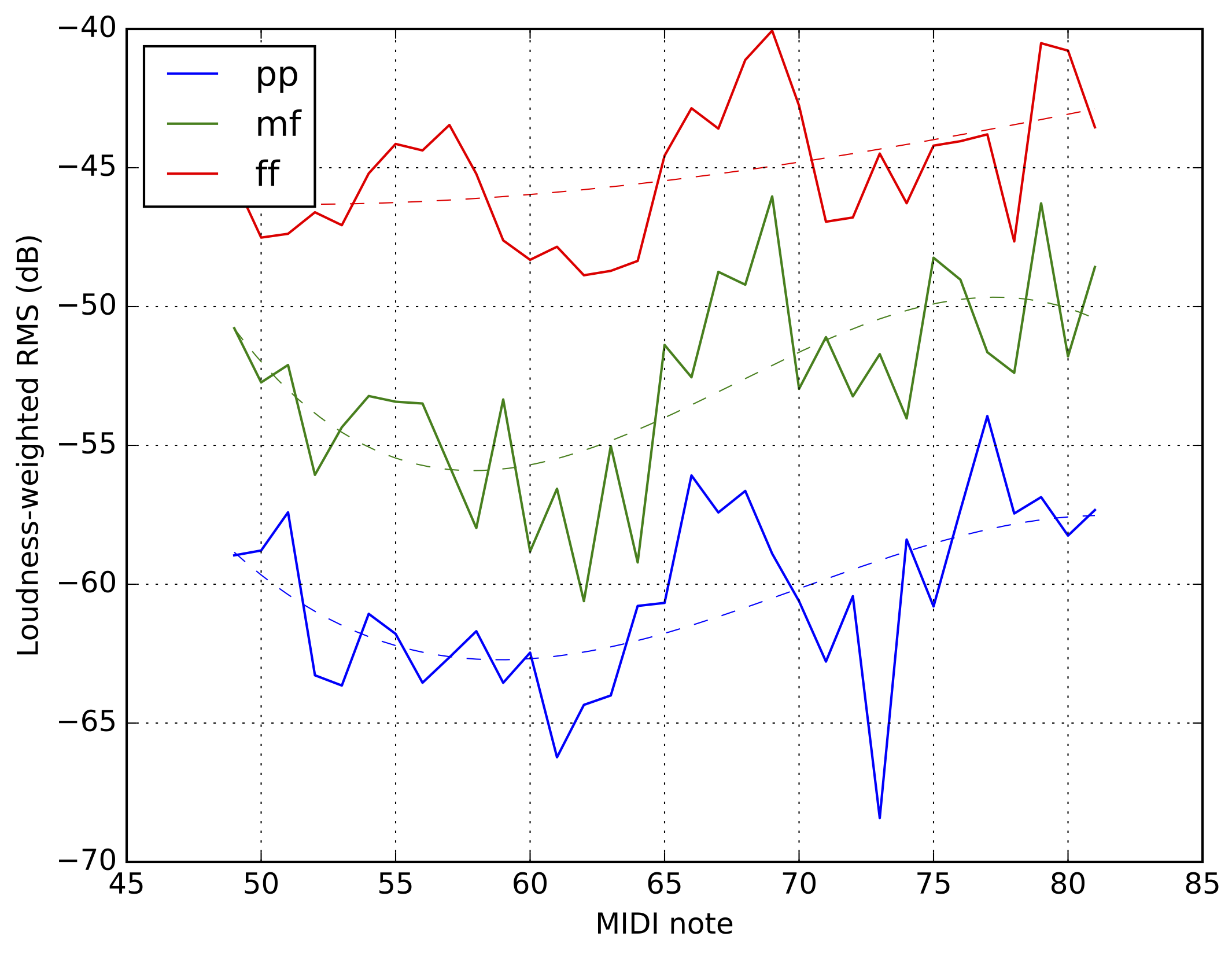}}\\
\caption{RMS volume curves of the Alto Saxophone ordinario notes at three different level of dynamics, in three different SOL-related datasets: (a) the original SOL; (b) the Orchids version of SOL; (c) our OrchideaSOL. The volume is calculated by windowing the signal and weighting each window's RMS according to a measure of loudness obtained via the \emph{pyloudnorm} library \cite{christian_steinmetz_2019_3551801}. Cubic least-square regressions are shown with dashed lines. \label{fig:volcompsax}}
\end{figure*}

\subsection{Selection}
\begin{sloppypar} 
We include 12253 sound files in the free OrchideaSOL dataset, amounting to 35\% of the total number of playing techniques within SOL.
The choice was motivated by the desire to provide an initial set of techniques rich enough to allow experimentation with the Orchidea tools.
This initial set includes: ordinario, sforzati, artificial harmonics, pizzicati, vibrati, slaps, aeolian sounds, brassy sounds, stopped sounds, flatterzunge, discolored fingerings, harmonic fingerings, notes played col legno (tratto or battuto), tremoli and bisbigliando, pedal tones, whistle tones, key clicks, jet whistles, and a few other extended techniques.
\end{sloppypar}

Due to the fact that currently Orchidea works with static targets, we exclude all techniques with time-varying dynamics, such as crescendo and descrescendo.

\subsection{Retuning}
A large part of samples in the original SOL distrubition were very audibly out of tune, therefore a retuning passage was inescapable. To avoid retuning all samples, we decided to only process those notes whose pitch error was above 10 cents and below 80 cents. The few samples with error above 80 cents were flagged and inspected manually on an individual basis (see section \ref{sec:manualcorr}).
The resampling was done via the Python \emph{resampy} module, with its \emph{kaiser\_best} filter\footnote{Resampy web page: \url{http://ccrma.stanford.edu/~jos/resample/}}. The fundamental frequency was estimated via Essentia's \emph{PitchYin} descriptor \cite{bogdanov2013essentia}, and verified by ear for every resampled sound.

\subsection{Volume compensation}
%
%

Although the third channel is by far the best channel to work with, as far as recording quality, it has the important drawback that its level and positioning had been often modified during the course of recording sessions.

A subset of recordings (Alto Sax, Accordion, Harp, Guitar and Bass Tuba) had precise reports for such modifications.
Most of the other instruments, unfortunately, did not.
In order to retrieve more natural relationships between the instrument dynamics, whenever we could, we reverted the volume modifications according to the reports; when we could not do that, we extrapolated the rationale behind them and we tried to infer volume adjustments family-wise, with the help of analysis of volume curves across the different levels of dynamics.
We initially tried to infer the loudness differences by comparing the signal of the third microphone with the standard stereo couple, but this proved to be more intricate, invasive and error-prone than an a-posteriori analysis of volume curves.

We decided to avoid any nonlinear signal processing, such as limiters or compressors.
Hence, we had to apply a global negative makeup on the dataset to avoid clipping; we recalibrated local make-up factors in order to account for macroscopic differences between families of instruments.
For example, in the original dataset, a flute \ff was much louder than a trumpet \ff.
Although we are aware that our choices are far from perfection --- whatever ``perfection'' may mean in a similar recovery task ---, we believe that the volume-compensated dataset is more faithful to the relationships between instrumental dynamics than the original one.
As such, we hope that it will prove more effective in orchestration tasks.

Figure \ref{fig:volcompsax} illustrates the application of our volume compensation procedure in the case of the Alto Saxophone. 
In the online repository of Orchidea, we provide a list of all the gain transformation which we applied.

\subsection{Resampling}
In SOL, some of the playing techniques were sampled by whole tones, minor thirds or major thirds.
Moreover, some of the samples were missing altogether.
Because Orchidea does not perform automatic pitch shifting of samples during the analysis process, we decided, for most of the playing techniques, to fill in the remaining notes by resampling the nearest ones, up to a tone upwards or downwards. 

\subsection{Renaming}
The naming in SOL was at times inconsistent and hard to parse properly.
Filenames usually included four fields (instrument, playing technique, pitch, dynamics), and possibly a fifth one (other specifications), all separated by dashes.
However, dashes were often used inside playing techniques themselves, as well as a description of pitch combinations, such as multiphonics of glissandi.
Moreover, in few instances, some of the fields were dropped as they were not applicable to the sample at hand.
In the release of OrchidaSOL, we devised a semi-automatic script to fix these issues and we performed the renaming of the sound files.\footnote{A renaming, slightly different from ours, was already accomplished in later version of SOL, such as the one used in Orchids.
However, these versions had the important drawback of having been normalized, with seemingly no trace of the original levels.}
The general filename has been brought to the standard form:
$$\texttt{<{\kk}i{\kk}n{\kk}s{\kk}t{\kk}r{\kk}>{\kk}-{\kk}<{\kk}p{\kk}s{\kk}>{\kk}-{\kk}<{\kk}p{\kk}i{\kk}t{\kk}c{\kk}h{\kk}>{\kk}-{\kk}<{\kk}d{\kk}y{\kk}n{\kk}>{\kk}-{\kk}<{\kk}o{\kk}t{\kk}h{\kk}e{\kk}r{\kk}>{\kk}-{\kk}<{\kk}r{\kk}e{\kk}s{\kk}>{\kk}.{\kk}w{\kk}a{\kk}v}$$
where \texttt{<ins>} is the abbreviated instrument name, \texttt{<ps>} is the playing technique, \texttt{<pitch>} is the pitch in textual form (under the convention that middle C is C4), \texttt{<dyn>} is the dynamics and \texttt{<other>} is any other meaningful specification (e.g. string number, alternative version number, etc.).
Any among \texttt{<pitch>}, \texttt{<dyn>} and \texttt{<other>} can be replaced by the letter `N' when such property is not applicable to the recorded file (e.g. pitch is irrelevant when playing a string instrument on the tuning pegs, and so on). All of the five property always appear in the file names (even though most of the time the \texttt{<other>} property is just `N'). If a mute is applied, the instrument name becomes \texttt{<instr>+<mute>}. 

\begin{sloppypar}
The \texttt{<res>} property includes information on the file resampling, namely: whether the tuning of the note was adjusted, or whether the note did not appear in the original dataset and was resampled from another one. The first information is in the form \texttt{T<amount><u|d>}, while the second is in the form \texttt{R<amount><u|d>}. The amount is always an integer number of cents; `u' stands for upwards and `d' stands for `downwards'. If no resampling was performed, the \texttt{<res>} property is `N'.
\end{sloppypar}

We rewrote the names of playing techniques so as to avoid any dash or non-ASCII character.
Furthermore, we harmonized discrepancies between some pairs of synonym categories, such as \emph{sforzato} vs. \emph{sforzando}.

\subsection{Manual corrections}\label{sec:manualcorr}
Finally, whenever we came across some file which was wrongly tagged or had evident issues in volume or pitch, we applied a manual correction.
The list of all the manual correction is provided in the Orchidea repository (see section \ref{sec:distrib}). 
Some issues were found in certain folders which could not be solved (e.g. missing or corrupted samples); a list of open issues is also provided within the repository.

\section{Baseline classification}
Together with the dataset, we decided to include some baseline classification results on three specific tasks: instrument recognition, playing technique recognition and note recognition. The tasks have different degrees of difficulty, given the unbalanced number of examples per class and the intrisic nature of each problem (instrument recognition has 32 classes, playing technique recognition has 89 classes and note recognition has 145 classes\footnote{This is due to the presence in the dataset of multi-pitch techniques, such as play-and-sing (playing a given note and singing another one).}).

The classification pipeline has been the same for each task and included, after a standardization phase, several classifiers: $K$-nearest neighbors (kNN), logistic regression (LogReg), support vector machines (SVC) and random forest with 10 estimators (RF10).
The sounds have been analysed using 20 MFFCs; the Python code used for the classification is included in the dataset distribution and can be referred for a detailed description of the parameters of each classifier.
Table \ref{tab:classif} details the results of our experiments (accuracy). 
In all cases, the random forest obtained the best results.
The playing technique recognition task appears to be the most difficult and the accuracies drop consequently.

\begin{table}[]
\begin{center}
\begin{tabular}{llll}
\hline
                                                              & Instrument & Playing technique & Note \\ \hline
kNN                                                           & .85        & .50           & .80  \\
LogReg                                                        & .85        & .58           & .87  \\
SVC                                                           & .92        & .73           & .87  \\ 
RF10                                                          & .95        & .90           & .90  \\\hline
\end{tabular}
\caption{Classification results (accuracy) for instrument recognition, playing technique recognition and note recognition. \label{tab:classif}}
\end{center}
\end{table}


It is important to remark that, in this context, we did not want to provide state-of-the-art classification results but we only wanted to give reference baselines for further experimentation.
We refer to \cite{lostanlen2018dlfm} on the potential of SOL and its derivatives for scientific research in machine listening.

\section{Distribution}\label{sec:distrib}
OrchideaSOL is distributed via the Ircam Forum\footnote{https://forum.ircam.fr/} and can be freely downloaded upon subscription.

\subsection{Distribution of OrchideaSOL metadata on Zenodo}

Besides its usage in musical creation, OrchideaSOL has the potential of advancing knowledge in scientific research.
Indeed, the wealth of playing techniques that is afforded by OrchideaSOL, as well as the consistency of its recording conditions, makes it an ideal test bed for timbre modeling.
In particular, samples in OrchideaSOL from different playing techniques are aligned in terms of onset time, fundamental frequency, and loudness level.
Thus, OrchideaSOL allows to devise systematic protocols on music perception and cognition in human subjects.
Furthermore, OrchideaSOL may be employed to train and evaluate machine listening software on various music information retrieval (MIR) tasks, including instrument classification, playing technique classification, and fundamental frequency estimation.

To facilitate the adoption of OrchideaSOL by the research community, we generate a metadata file in CSV format which summarizes the attributes of every audio sample.
This metadata file expedites the need for writing a custom parser on the part of the MIR practitioner \footnote{The metadata of OrchideaSOL can be downloaded at:\\ \url{https://doi.org/10.5281/zenodo.3686251}}.

For the sake of research reproducibility, we provide an official split of OrchideaSOL into five non-overlapping folds, as an additional column to the metadata spreadsheet.
Here, it is crucial that all folds have the same distribution of labels.
To achieve this goal, we apply an algorithm named Entrofy \cite{huppenkothen2019entrofy}, originally designed for cohort selection among human subjects.
After convergence, we verify that all folds fare equally in terms of instruments, playing techniques, pitch range, and intensity dynamics.

\subsection{Distribution of TinySOL in the mirdata package}

Furthermore, we provide a Python module allowing every user of TinySOL to guarantee that they have access to the dataset in its pristine form.
This module is integrated into the \emph{mirdata} package \cite{bittner2019ismir}, an open-source initiative for the reproducible usage of datasets\footnote{The mirdata Python package can be installed with the command:\\ \texttt{pip install mirdata}.}.
The key idea behind \emph{mirdata} is for the dataset curators to upload a dataset ``index'', in JSON format, which summarizes the list of files as well as their MD5 checksums.

To this end, we upload a copy of TinySOL to the Zenodo repository of open-access data\footnote{The audio and metadata of FullSOL can be downloaded at:\\ \url{https://doi.org/10.5281/zenodo.3632192}}.
Because Zenodo is developed by the European OpenAIRE program and operated by CERN, it has a anticipated lifespan of multiple decades.
In addition, the presence of TinySOL as an unalterable dataset on Zenodo associates it to a digital object identifier (DOI), which is directly citable in scientific publications.
For this reason, our implementation of the TinySOL module for \emph{mirdata} points to the Zenodo repository as a reliable source.

A track from TinySOL may be loaded as follows:
\begin{lstlisting}[
    language=Python,
    basicstyle=\footnotesize\ttfamily]
from mirdata import tinysol
data_home = "mir_datasets/TinySOL"
tinysol.download(data_home=data_home)
dataset = tinysol.load()
track = dataset["Fl-ord-C4-mf-N-T14d"]
\end{lstlisting}
The corresponding waveform can be loaded (via librosa \cite{mcfee2020librosa}) by accessing the \texttt{track.audio} property.
Likewise, metadata for this track corresponds to other properties such as \texttt{instrument\_abbr}, \texttt{technique\_abbr}, \texttt{dynamics}, and \texttt{pitch\_id}.

Because \emph{mirdata} is version-controlled and released under a free license, it protects TinySOL against the eventuality of slight alteration, either accidental or deliberate.
Indeed, the function \texttt{mirdata.tinysol.validate()}
 compares all MD5 checksums in the local repository against the public checksums.
In case of mismatch, \emph{mirdata} lists all files which are missing, spurious, or corrupted.
Therefore, this function offers a guarantee to researchers that they are using the pristine version of TinySOL.

\subsection{Distribution of pre-computed features}
In addition to raw audio samples, the OrchideaSOL repository includes:

\begin{itemize}
\item Statistics on the number of samples (also see Figures \ref{fig:countinstr} and \ref{fig:countps}).

\begin{figure}[h]
\centering
\includegraphics[width=\columnwidth]{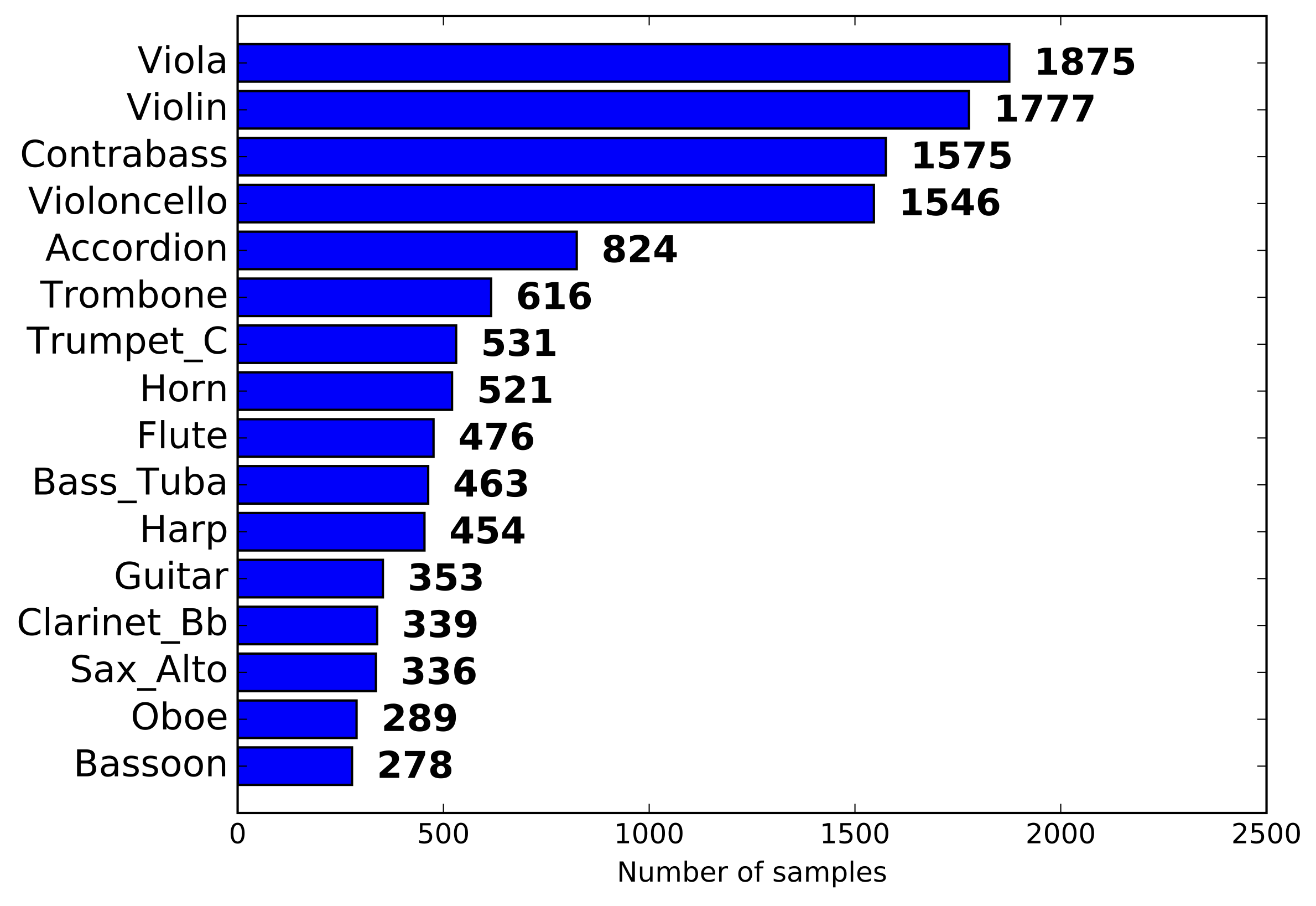}
\caption{Distributions of samples by instrument in the OrchideaSOL dataset (samples with mutes are included in the count) \label{fig:countinstr}}
\end{figure}

\begin{figure}[h]
\centering
\includegraphics[width=\columnwidth]{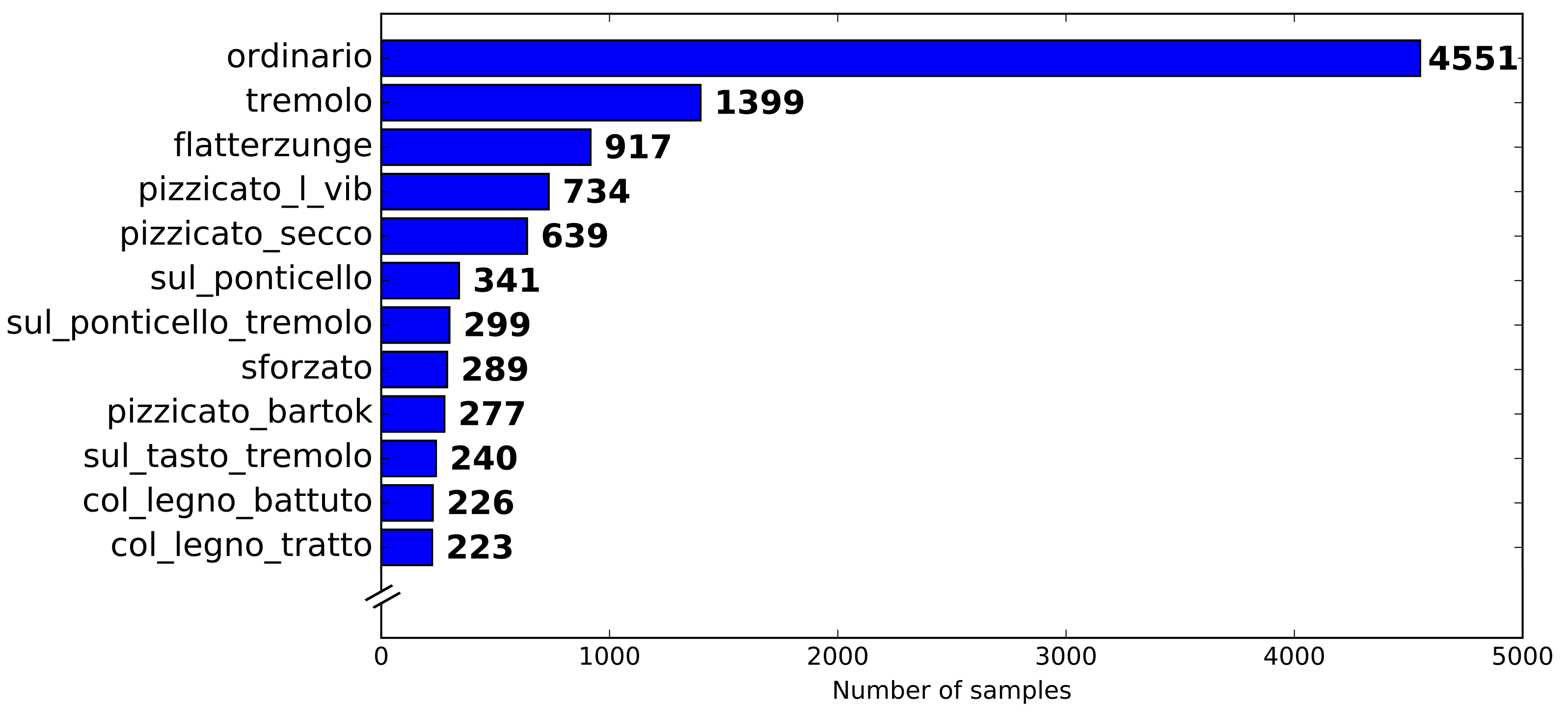}
\caption{Distributions of samples by playing technique in the OrchideaSOL dataset. Only playing techniques with more than 200 entries are displayed (12 out of 52). \label{fig:countps}}
\end{figure}

\item Volume curves for each instrument and playing techniques, with plotted cubic least-squares regression. 
For all the ``ordinario'' samples, we also provide the coefficient of the quadratic two-dimensional regression formula for the loudness-weighted RMS ($L_{\textrm{dB}}$):
\begin{equation*}
L_{\textrm{dB}} = \alpha + \beta M + \gamma D + \delta MD + \phi M^2 + \psi D^2  
\end{equation*}
where $M$ is the MIDI number and $D$ is a numeric value assigned to each dynamic marking according to the convention
\begin{multline*}
\ldots, \quad \pp = -2.5, \quad \p = -1.5, \quad \mezzopiano=0.5,\\
\mf=0.5, \quad \f = 1.5, \quad \ff = 2.5, \quad \ldots
\end{multline*}

Figure \ref{fig:flute} show some examples of volume curves for the flute (ordinario).
The quadratic regression for its loudness-weighted RMS is 
\begin{multline*}
L_{\textrm{dB}, \textrm{flute}} \approx -82.3301 + 0.2732 M + 3.0831 D + \\ + 0.0194 MD + 0.0008M^2 + 0.0823 D^2
\end{multline*}
These formulas can be also used to roughly model instrumental dynamics in other contexts, such as synthesis or sampling. 

\begin{figure*}[h]
\centering
\subfloat{%
      \includegraphics[width=0.49\textwidth]{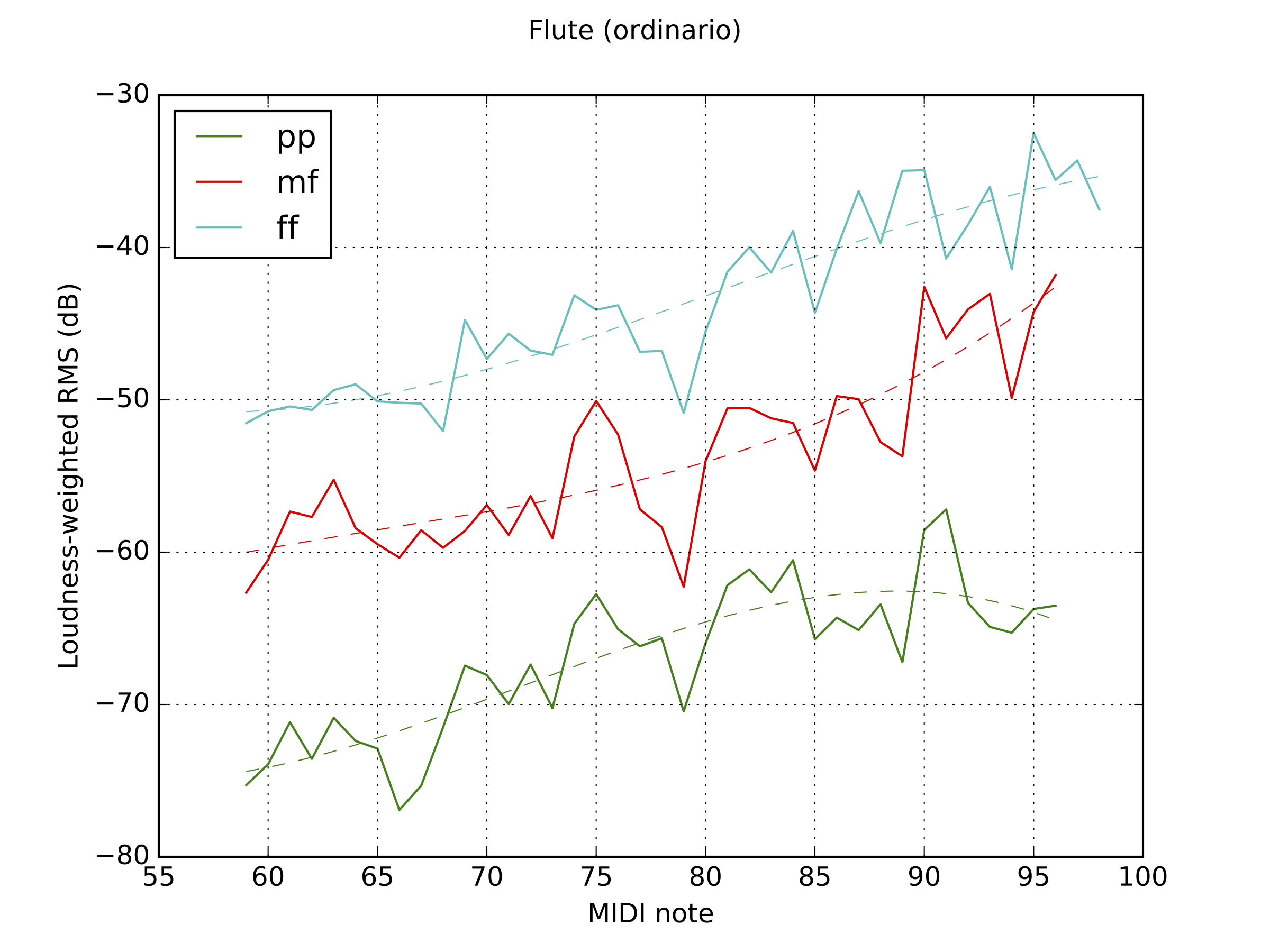}}
\hspace{\fill}
   \subfloat{%
      \includegraphics[width=0.49\textwidth]{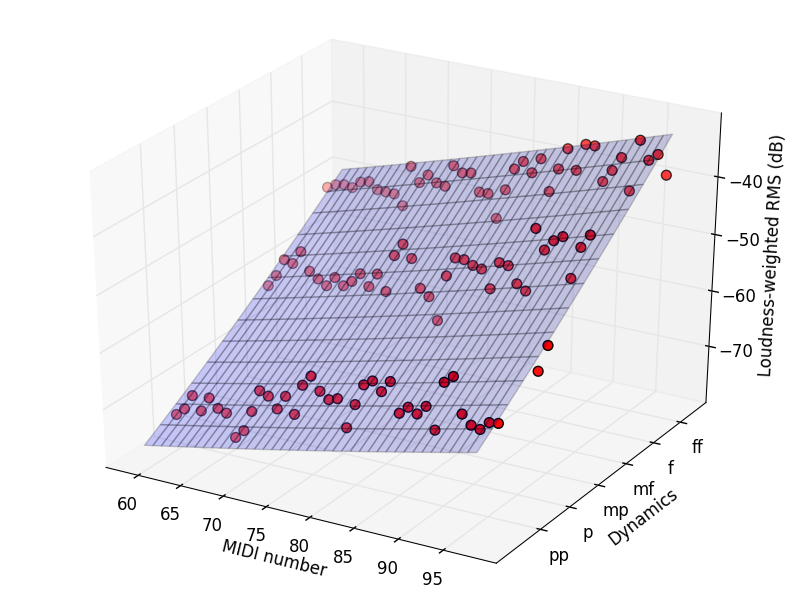}}
\caption{Volume curves for the flute (left, cubic regression lines are dashed) and three-dimensional representation of the corresponding regression surface (right). \label{fig:flute}}
\end{figure*}


\item Spectral analysis: for each sample, we provide the first 1024 amplitude bins of its average spectrum (analyzed with window size: 4096 samples, hop size: 2048 samples);

\item MFCC analysis: for each sample, we provide the average of the first 20 MFCCs (analyzed with window size: 4096 samples, hop size: 2048 samples).

\item Spectral envelope: for each sample, we provide the first 1024 bins of the average spectral envelope (analyzed with window size: 4096 samples, hop size: 2048 samples); the spectral envelope is computed by applying a lowpass window of 80 coefficients (liftering) to the cepstrum and by taking again the Fourier transform; briefly, given the signal S:
\begin{equation*}
	E = FFT (W_{LP} (FFT^{-1} (\log (|FFT (S)|)).
\end{equation*}
\item Spectral peaks: for each sample, we provide the average of the first 120 peaks of the magnitude spectrum  analyzed with window size: equal to 4096 samples and hop size equal to 2048 samples.

\item Spectral moments: for each sample, we provide the average of the first 4 spectral moments \cite{peeters2004large}: centroid, spread, skewness and kurtosis, analyzed with window size equal to 4096 samples and hop size equal to 2048 samples).

\end{itemize}

\section{Conclusion}
We have introduced OrchideaSOL, a free subset of the SOL dataset modified and tailored to be reliable in target-based computer-aided composition tasks.
We expect the generated scores to be more faithful in real-life orchestral scenarios than with the previous SOL dataset.
Moreover, we believe that OrchideaSOL has a strong potential for advancing scientific research in music cognition, music information retrieval, and generative models for sound synthesis.


\begin{acknowledgments}
We would like to thank Hugues Vinet, Jean-Louis Giavitto and Gregoire Beller for sharing with us the collection of internal Ircam reports on the SOL recordings and for agreeing to release OrchideaSOL for free.
\end{acknowledgments} 

\bibliography{SOLicmc2020}

\end{document}